\documentclass[reprint,preprintnumbers,amsmath,amssymb,nopacs]{revtex4}
\usepackage{graphicx}
\usepackage{xcolor}

\begin{document}
\title{Random walks on uniform and non-uniform combs and brushes}

 \author{Alex V. Plyukhin} 
\email{aplyukhin@anselm.edu} \affiliation{
  Department of Mathematics, Saint Anselm College, Manchester, NH,
  USA} 
\author{Dan Plyukhin} 
\email{dplyukhin@cs.toronto.edu} \affiliation{ Department of
  Computer Science, University of Toronto, Toronto, ON, Canada}

\date{\today}

\begin{abstract}
We consider random walks on comb- and brush-like graphs
consisting of a base (of fractal dimension $D$) decorated with attached side-groups.  
The graphs are also characterized by the fractal dimension $D_a$ of a set of anchor points where side-groups are attached to the base.
Two types of graphs are considered. Graphs of the first type are uniform in the sense that 
anchor points are distributed periodically over the base, and thus form a subset of the base with dimension $D_a=D$. Graphs of the second type are decorated with side-groups in a regular yet non-uniform way: the set of anchor points has fractal dimension smaller than that of the base, $D_a<D$.  For uniform graphs, a qualitative method for evaluating the sub-diffusion exponent suggested by Forte et al.~for combs ($D=1$) is extended for brushes ($D>1$) and numerically tested for the Sierpinski brush (with the base and anchor set built on the same Sierpinski gasket). 
As an example of nonuniform graphs we consider the Cantor comb composed of a  one-dimensional base and side-groups, the latter attached to the former at anchor points forming the Cantor set. 
A peculiar feature of this and other nonuniform systems
is a long-lived regime of super-diffusive transport when side-groups are of a finite size.

\end{abstract}

%\pacs{05.40.Fb, 05.45.Df, 02.50.Ey}

\maketitle

\section{Introduction}

Comb- and brush-like graphs and networks are branched systems 
consisting of a base of fractal dimension $D$ and a collection of
identical side-groups, each of fractal dimension $d$, attached to the base according to a certain protocol.
Structures with a one-dimensional base $D=1$ are often referred to as combs (see Fig.~1)
and those with $D>1$ as brushes (see Fig.~2); the unifying term ``bundled structures'' was coined  for both graph types in Ref.~\cite{CR}. 

In this paper we will find it instructive to further classify bundled structures by 
the fractal dimension $D_a$ of the anchor set, i.e. the subset of the base consisting of the points (called anchor points) at which the side-groups are attached; see Fig.~1. 
We shall say that a bundled structure is uniform (or uniformly decorated) if side-groups are attached to the base in a periodic manner, so that $D_a=D$; see Figs.~1 and 2. Otherwise, 
if $D_a<D$, the structure will be called non-uniform. 
An example of a non-uniform bundled structure is the Cantor comb ($D=d=1$, $D_a=\ln 2/\ln 3$); 
see Fig.~6.  We will show that uniform and non-uniform bundled structures may differ considerably  in their transport properties, as well as in the methods for their analysis.

%\enlargethispage{1\baselineskip}

Random walks on uniform comb-like structures were first studied as a simplified model of transport on percolation clusters~\cite{Havlin_paper,Havlin_book}, but 
in recent years the problem has also been addressed from different perspectives, and found to be relevant to many other natural and man-made systems and processes 
(e.g. transport in branched polymers, porous media, spiny dendrites of neuron cells, and artificial networks); see~\cite{CR,Mendez,Mendez2,Baskin,Forte,Iomin,Agliari} and references therein. 
Typically, transport characteristics (like the random walk dimension) of the base or a side-group \emph{in isolation} are already known, and one is interested in finding those characteristics
for the corresponding composition.
Of particular interest is the exponent $\alpha$, which describes the mean-square displacement along the base $ \langle r^2(t)\rangle\sim t^\alpha$. For bundled systems with infinitely extended side-groups,  
transport along the base usually exhibits 
sub-linear ($\alpha<1$ or even slower, e.g. logarithmic) time dependence.
For example, for a simple uniform comb composed of an unbounded one-dimensional base and side-groups ($D=D_a=d=1$), transport along the base is sub-diffusive with $\alpha=1/2$. 
Diffusion along the base of a uniform brush, composed of a two-dimensional base ($D=D_a=2$) and one-dimensional side-groups ($d=1$), is governed by the same transport exponent $\alpha=1/2$.
While these results for $\alpha$
can be obtained theoretically in a simple manner~\cite{Havlin_paper,Havlin_book,Baskin}, its evaluation 
for more complicated structures 
%even when analytically feasible 
may be quite involved~\cite{CR,Iomin}, particularly for those with non-trivial 
fractal dimensions $D$, $D_a$, $d$.
On the other hand, for the special case of uniform comb-like structures ($D=D_a=1$), 
Forte et al. recently suggested an attractive qualitative method to determine
the transport exponent $\alpha$ without detailed combinatorial calculations, 
based on a simple matching argument~\cite{Forte}.

%\enlargethispage{2\baselineskip}

The first goal of this paper is to 
extend 
%(in a quite straightforward manner)  
the qualitative approach by Forte et al.
to uniform {\it brush}-like structures with a fractal base dimension, $D=D_a>1$; this is quite straightforward and will be discussed in Section II, with support in Section III by numerical simulations on the uniform  Sierpinski brush ($D=D_a=\ln 3/\ln 2$ and $d=1$). 
Simulation of transport in bundled structures, 
characterized by non-integer dimensions, has some interesting peculiarities which
%to the best of our knowledge
we feel have not been fully addressed in 
the literature so far.

The second  goal of the paper is to
simulate random walks on non-uniform bundled structures, like the Cantor comb (see Fig.~6), for which $D_a<D$. 
In Section IV we confirm a theoretical result for 
the transport exponent $\alpha$ suggested for such systems in
Ref.~\cite{Iomin}. 
We also studied non-uniform combs with \emph{finite} side-groups,
observing for that case a very long transient regime of 
super-diffusive transport, with mean-square displacement increasing faster than linearly with time. We show that the long duration of this transient makes the heuristic method of Forte et al. inapplicable to non-uniform bundled structures.

\section{Uniform systems: Matching argument}
Let us recapitulate the matching 
argument of Forte et al.~\cite{Forte}. 
Consider a  generic uniform comb ($D=D_a=1$) decorated 
with identical side-groups, each characterized by fractal dimension $d$, random walk dimension $d_w$, and spectral dimension 
$d_s=2d/d_w\le 2$; see Fig.~1.
Assume that the base and side-groups
are discrete and that random walks are characterized by a single
hopping rate $w$ and a hopping distance of unit length. 
Diffusion on the base in isolation (ignoring side-groups) is normal, whereas diffusion 
on an isolated side-group (detached from the base) may be either normal or anomalous. 
The corresponding mean-square displacements as functions of time are
\begin{eqnarray} 
\langle x^2(t)\rangle_b\sim w\,t,\qquad
\langle r^2(t)\rangle_{sg}\sim (w\,t)^{2/d_w},
\label{isolated}
\end{eqnarray}
where subscripts $b$ and $sg$ refer to the isolated base and an isolated side-group, respectively.
Given (\ref{isolated}), one  wishes to evaluate the exponent $\alpha$ for the mean-square displacement
\begin{eqnarray}
\langle x^2(t)\rangle\sim (w\,t)^\alpha
\end{eqnarray}
along the base of the comb when decorated with infinitely extended side-groups.

 \begin{figure*}[t]
\includegraphics[height=6.4cm]{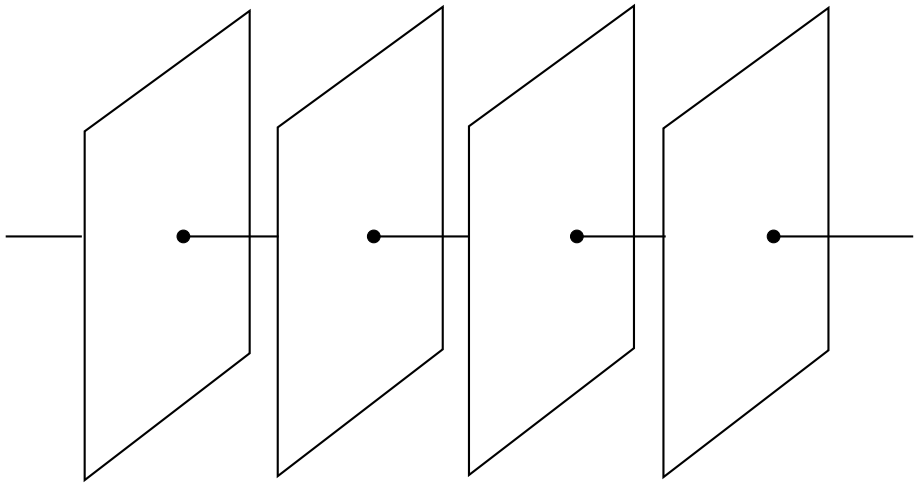}
\caption{A cartoon of a fragment of a uniform comb structure. 
The base is one-dimensional ($D=1$), while side-groups,
depicted here as planes, are actually fractals of dimensions $d\ge 1$
and spectral dimension $d_s=2d/d_w\le 2$. The anchor points, where
side-groups are attached to the base, are denoted
by the symbol $\bullet$. For uniform systems like this, the set of anchor points has dimension $D_a$ the same as for the base, $D_a=D$. For an example of non-uniform systems with $D_a<D$, see Fig.~6.
}
\end{figure*}

To this end, consider first a comb composed of the infinite base decorated with
{\it finite} side-groups of linear size $L$. For such a system,
on sufficiently long time scales
$t\ge t_c(L)$
the diffusion along the base is expected to be  
normal, but with an effective $L$-dependent hopping rate $W(L)<w$,
\begin{eqnarray} 
\langle x^2(t)\rangle\sim W(L)\,t, \qquad t\ge t_c(L).
\label{long_time}
\end{eqnarray}
The characteristic time  $t_c(L)$ can be estimated as the one needed
for a random walker to diffuse along an  isolated side-group 
to the distance of order $L$, 
$\langle r^2(t_c)\rangle_{sg}\sim (w\,t_c)^{2/d_w}\sim L^2$, which gives
\begin{eqnarray} 
t_c(L)\sim w^{-1} L^{d_w}.
\label{tc}
\end{eqnarray}
As for the effective hopping rate $W(L)$, 
one expects it to be 
proportional to the occupation probability of the anchor point
for random walks on an isolated side-group of size $L$. The assumed condition on the spectral dimension 
$d_s=2d/d_w\le 2$ implies that diffusion on an isolated side-group is compact~\cite{Havlin_book}, 
so one can evaluate the effective rate on a time scale $t>t_c(L)$ as
\begin{eqnarray} 
W(L)\sim w\,L^{-d}.
\label{W}
\end{eqnarray}
Eqs.~(\ref{long_time})-(\ref{W}) completely specify diffusion on  the long time scale, $t>t_c(L)$.
%This estimation assumes that the spectral 
%dimension $\tilde d=2d/d_w$ of a sidegroup is less than two,
%so that  random walks on  the sidegroup is compact: given the 
%mean-square diaplacement $L^2$, the probability 
%distribution is approximately uniform in the shpere of radius $L$ 
%about the starting point.

On time-scales shorter than $t_c(L)$, 
%(but longer than $w^{-1}$) 
a random walker would not feel the boundary of the side-groups, and the mean-square
displacement along the base is expected to have the same form as in
the case of unbounded side-groups,
\begin{eqnarray}
\langle x^2(t)\rangle\sim (w\,t)^\alpha, \qquad t< t_c(L).
\label{short_time}
\end{eqnarray} 
Assuming that the transition from the short-time (\ref{short_time}) to 
long-time (\ref{long_time}) behaviours is sufficiently abrupt,
%(occurs during a time interval $\Delta t$ which weakly depends on $L$)
one expects that 
the two expressions must approximately match at $t=t_c(L)$, which gives
\begin{eqnarray}
W(L)\,t_c(L)=[w\,t_c(L)]^\alpha.
\label{match1}
\end{eqnarray} 
Substituting expression (\ref{tc}) for $t_c(L)$ and (\ref{W}) for $W(L)$, 
one finds~\cite{Forte}
\begin{eqnarray}
\alpha=1-\frac{d}{d_w}=1-\frac{d_s}{2}.
\label{alpha_comb}
\end{eqnarray} 
This result coincides with that  of the rigorous  combinatorial approach~\cite{CR}
and is also in accordance with analytical and numerical studies
of specific comb-like structures reported in the 
literature~\cite{Havlin_book,Havlin_paper,Mendez,Baskin}.

Our first goal 
is to extend the above reasoning for brush-like systems
with base dimension $D>1$. As before, we shall assume that 
the base is decorated with side-groups uniformly, i.e.
$D_a=D$, see Fig.~2. For brush-like systems,  diffusion may be anomalous 
not only on isolated side-groups but also on the isolated base, 
\begin{eqnarray} 
\langle r^2(t)\rangle_b\sim (w\,t)^{2/D_w},\qquad
\langle r^2(t)\rangle_{sg}\sim (w\,t)^{2/d_w}.
\label{isolated2}
\end{eqnarray}
Here $D_w$ and $d_w$ are random walk dimensions for an isolated base
and side-group, respectively. 
Consider first  a brush with finite side-groups of 
linear size $L$.  As for combs with finite side-groups, we expect that 
on a sufficiently long time scale $t>t_c(L)$ 
diffusion along the base is of the same type, 
i.e. has the same 
random walk dimension $D_w$ as for 
an isolated base, but with a re-normalized $L$-dependent hopping rate $W(L)$,
\begin{eqnarray} 
\langle r^2(t)\rangle\sim [W(L)\,t]^{2/D_w}, \qquad t\ge t_c(L),
\label{long_time2}
\end{eqnarray}
while for $t<t_c(L)$ the mean-square displacement follows the same law 
$\langle r^2(t)\rangle\sim (w\,t)^\alpha$ as for the infinite system.  
Assuming that the transition between the two diffusion regimes
occurs sufficiently fast, one expects the corresponding expressions
for the mean-square displacement to be  approximately 
equal  at 
the transition time $t=t_c(L)$, 
\begin{eqnarray} 
[W(L)\,t_c(L)]^{2/D_w}=[w\,t_c(L)]^\alpha.
\label{match2}
\end{eqnarray}
This matching relation is analogous to that for combs, 
Eq. (\ref{match1}). 
%Simulation results of the next section show that this relation should actually be understood in the limit of large $L$. 
%however this is unimportant for the purpose of 
%the present derivation.

The next assumption is that the effective rate 
$W(L)$ and the crossover time $t_c(L)$ depend on  the structure of
side-groups but not on that  of the base, and therefore estimations (\ref{tc}) and (\ref{W}) for those quantities obtained above for combs  
should be  valid for brush-like structures also. Then substituting 
(\ref{tc}) and (\ref{W}) into (\ref{match2}), one obtains
\begin{eqnarray} 
\alpha= \frac{2}{D_w}\,\left(
1-\frac{d}{d_w}
\right)=\frac{D_s}{D}\,\left(
1-\frac{d_s}{2}
\right),
\label{alpha_brush}
\end{eqnarray}
where $D_s=2D/D_w$ and $d_s=2d/d_w$ are spectral dimensions of the isolated base and side-group, respectively.
This result is in agreement with the combinatorial evaluation of Ref.~\cite{CR}, and in the next section we shall verify it with  numerical simulations
for a specific brush structure with one-dimensional side-groups ($d=1$, $d_w=2$). 
For the latter case, the expression (\ref{alpha_brush}) takes 
the  simple form
$\alpha=1/D_w$.

As the structure of (\ref{alpha_brush}) suggests, that expression holds
for $d_s<2$,  which is the  condition of compactness of  diffusion on isolated side-groups. Also, recall that both expressions (\ref{alpha_comb}) for combs and  (\ref{alpha_brush}) for brushes hold for uniform systems only, 
assuming the  condition $D_a=D$.
We postpone the discussion of  non-uniform structures with $D_a<D$ until Section IV.

 \begin{figure*}[t]
\includegraphics[height=6.6cm]{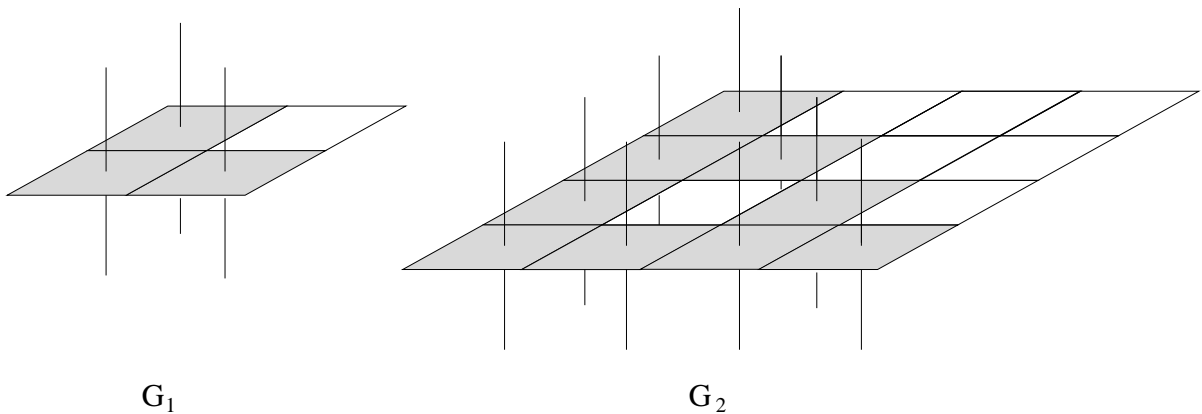}
\caption{
The first two generations of the uniform Sierpinski brush. The base is formed by (shaded) cells of the Sierpinski gasket of dimension $D=\ln 3/\ln 2$. One-dimensional side-groups attached to each cell of the base, so that 
the dimension of the set of anchor points $D_a$ is the same  as that of the base, $D_a=D=\ln 3/\ln 2.$}
\end{figure*}

\section{Simulation: Sierpinski brush}
Consider  a uniform brush with one-dimensional side-groups and 
a Sierpinski gasket~\cite{Havlin_book}  as a base; see Fig.~2. 
For this bundled structure, which we call the Sierpinski brush, the set of relevant dimensions is
\begin{eqnarray}
d=1,\qquad d_w=2, \qquad D=D_a=\ln 3/\ln 2, \qquad D_w=\ln 5/\ln 2,
\end{eqnarray}
and according to result (\ref{alpha_brush}) of the previous section, the mean-square displacement
along the base is expected to follow the sub-diffusive law 
$\langle r^2(t)\rangle\sim t^\alpha$ with 
$\alpha=1/D_w=\ln 2/\ln 5\approx 0.43$.
Numerical simulation results, presented in Fig.~4, support this prediction. In addition, they also show that 
the convergence to the above asymptotic dependence 
occurs very slowly, taking about $10^3$ steps.
Below we discuss some details of the simulation which, due to the fractal geometry of the base, are of interest on their own.

Numerical simulation of random walks on fractals 
typically involves the following two points. 
Firstly, in order to minimize finite size effects one needs
to run random walks on a fractal set of such a large size that 
explicitly storing the coordinates of each site in computer memory 
(say, in a multidimensional array) is infeasible.
Since a coordinate system is prerequisite to knowing how to
navigate the fractal,
one needs an effective (and efficient) algorithm for enumerating the
hopping sites and determining the labels of their neighbors.   
%This problem was first addressed for Sierpinski triangle in~\cite{Rammal}
%and in many papers later.
Secondly, once a pertinent labeling algorithm for a fractal set 
is developed, it may still be a nontrivial problem to express with it the
metric properties of the set. This second problem 
%(given a site's label or address, find the site's (Cartesian) coordinates)  
might be avoided if one is interested in such properties as the probability of 
return to the origin, but becomes inevitable
when one wishes to directly evaluate the mean-square displacement,
drift in an external field, etc.
As shown below (see also Ref.~\cite{Plyukhin}), for the Sierpinski brush
and similar systems both difficulties can be readily
resolved by embedding the fractal in a 2-dimensional Euclidean lattice
and writing the Cartesian coordinates of fractal sites as binary numbers.

Unlike the usual definition of the Sierpinski gasket, which is
defined ``from the outside, in'' we construct the base of the Sierpinski brush ``from the inside, out'' as the limit
of a sequence of inductively defined 
generations (see Fig.~3):  
The first generation $G_1$
is composed of three quarters of a square,
and specifies the pattern for higher generations;
each subsequent generation $G_{n+1}$ is defined
according to this pattern as three
copies of $G_n$.
For the purposes of simulation it suffices to let the base be a
generation of sufficient size, so that finite size-effects are negligible;
The presented results use $G_{30}$, with approximately $10^{14}$ jumping sites.
%Designed in this way, the base is a sublattice of a two-dimensional %lattice and has fractal dimension $D=\ln 3/\ln2$.
Note that we identify the jumping sites with the cells of the base 
(depicted as shaded squares in Figs.~2 and 3),
rather than the base's vertices.
The brush construction is finalised by
attaching one-dimensional discrete side-groups to  
each cell of the base.

As mentioned above, in order to identify a cell from the base
in the embedding 2D lattice, it is convenient to 
enumerate the cells of the lattice by pairs of Cartesian coordinates $(x,y)$, each
expressed in binary form. In $G_k$, the binary 
coordinates of each cell consist $k$ bits, i.e. digits, which are each either zero or one:
\begin{eqnarray}
x=(a_1 a_2 \cdots a_k),\qquad y=(b_1 b_2\cdots b_k), \qquad a_i,b_i\in \{0,1\}.
\label{address}
\end{eqnarray}  
As seen from Fig.~3,
the cells belonging to the base (i.e. the Sierpinski gasket cells) are those, and only those, cells of the $2D$ lattice 
for which the sum of bits in every position of the binary address
is either zero or one but not two:
\begin{eqnarray}
a_i+b_i<2, \quad {\mbox{for}} \quad i=1,2,\dots, k.
\label{condition}
\end{eqnarray}  
For example, in the base of second generation $G_2$ (see Fig.~3)
the cell with Cartesian
coordinates $x=3$, $y=1$ has the binary address $x=(11), y=(01)$. 
This cell is not 
in the base, and condition (\ref{condition}) is not satisfied for the second position, 
$a_2+b_2=2$. On the other hand, the cell with Cartesian coordinates
$x=2, y=1$ and the binary address $x=(10)$, $y=(01)$
belongs to the base, and condition (\ref{condition}) is satisfied 
for both positions,
$a_1+b_1=a_2+b_2=1<2$.

 \begin{figure*}[t]
\includegraphics[height=6.6cm]{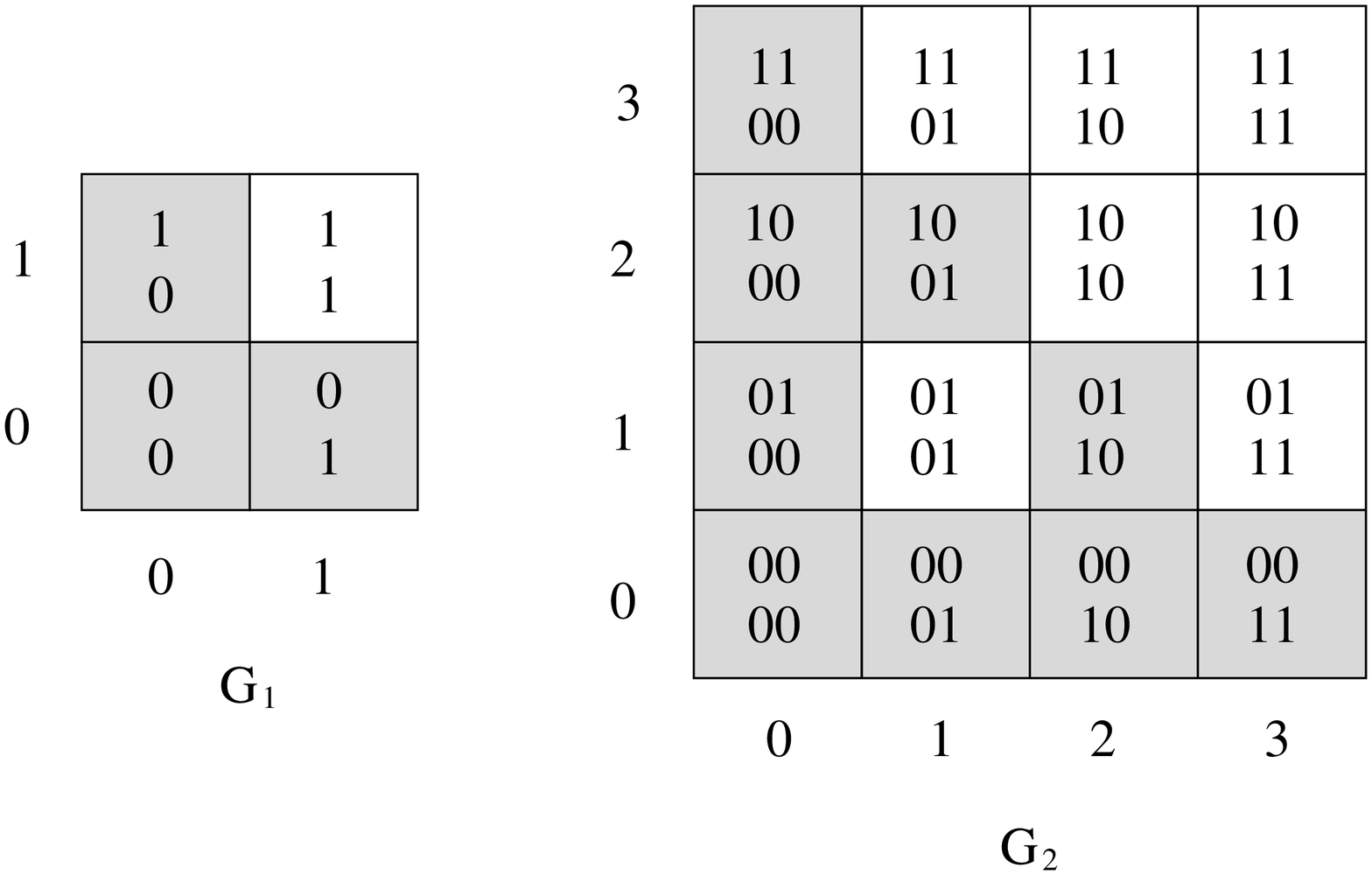}
\caption{An inductive composition and enumeration scheme for the   
first two generations of the Sierpinski gasket. Each cell of 
the embedding $2D$ lattice is labelled with its binary address, that is, 
the cell's Cartesian coordinates $x$ (bottom) and  $y$ (top) 
expressed in base two. For cells of the gasket the sum of
digits (bits) in every position is less than two.
}
\end{figure*}

Consider first {\it random walks on the isolated base}, i.e. 
on the Sierpinski gasket without side-groups.
%of generation $G_k$ with $k\gg 1$.
An initial site 
is selected by randomly generating a binary address
of the form (\ref{address}) subject to 
condition (\ref{condition}).
A particle is placed on the initial site and performs random walks according to the following two step protocol: 
First, the particle makes a virtual jump to 
a randomly selected nearest neighbour cell on the embedding $2D$ lattice.
In our simulation the coordination number  
of the embedding  
lattice  is $n_c=8$, so that the particle 
chooses  randomly between eight neighboring sites.
Second, we check that the new site satisfies (\ref{condition}) to determine
if the new site belongs to the base, i.e. if it is a cell of the Sierpinski gasket.
If the test is passed, the virtual jump is accepted. 
Otherwise (the new site does not belong to the base), the virtual jump is rejected,
and the particle remains on the original site. 
%This algorithm is rather time consuming since one needs to evaluate both %Cartesian and binary address of every tried site, yet it 
%easily solves both problems
%of finding a neighbour for a given cell of the gasket and of evaluation of the 
%mean-square displacement of the particle.

Depending on how one handles rejected jumps, 
the above protocol can be performed 
in two ways.
In the so-called ``blind ant scenario'', rejected jumps are counted as steps with zero
displacement but have the same duration as accepted jumps.
On the other hand, in the ``myopic ant scenario'' rejected jumps are totally virtual 
and not counted at all, i.e. have zero duration. In other words, a myopic ant
selects a new site to jump
only among those nearest neighbour sites which belong to the gasket.
%Our simulation confirm a common believed that for both blind and myopic 
%ant scenarios the  the mean-square displacement 
%$\langle x(t)\rangle \sim t^{2/D_w}$ is characterized by the same
%dimension $D_w$~\cite{Havlin_book,Rammal}. 
For both scenarios we found the random walk dimension to be the same, and  very close to 
$D_w=\ln 5/\ln 2\approx 2.32$.  This value is 
the same as  
for the more familiar triangular Sierpinski
gasket~\cite{Havlin_book}.
The equality of $D_w$ for the triangular and rectangular versions of the Sierpinski 
gasket, though intuitively expected, is perhaps not quite obvious considering that the two
fractals are not completely equivalent: for a triangular Sierpinski gasket
the coordination number for all cells (except the three apex cells) 
is $3$, while for our rectangular gasket this number is either $3$ or $4$. 
In our simulation this difference was found to  have no effect 
on the asymptotic parameters of random walks.
%We leave a theoretical justification of this as an %exercise for future studies.

%Can we apply the same exact scaling method developed for the triangular
%gasket for our rectangular gasket?

 \begin{figure}
  \includegraphics[height=6.2cm]{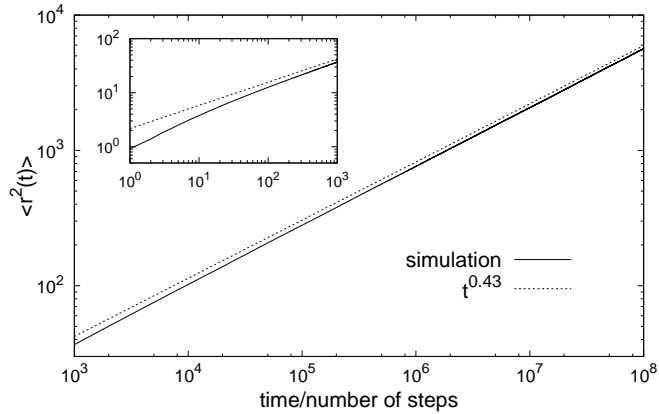}
  \caption{A log-log plot of the mean-square displacement 
  along the base of the Sierpinski brush 
(see Fig.~2) of generation $G_{30}$  with infinite side-groups 
at long times (and at shorter times in the inset). 
Random walks on the base are executed according to the myopic ant scenario.
Solid lines show the simulation results averaged over about $10^4$
 walks, and dashed lines correspond to the 
 theoretically expected dependence$\langle r^2(t)\rangle\sim t^{\alpha}$
with the exponent given by Eq.(\ref{alpha_brush}), $\alpha=1/D_w=\ln 2/\ln 5\approx 0.43$.}
\end{figure}

The extension of the above simulation scheme for the Sierpinski brush,
i.e. when one-dimensional side-groups are attached to every cell of the Sierpinski gasket (see Fig.~2) is straightforward.
The results presented in Fig.~4 confirm  theoretical prediction (\ref{alpha_brush}) for brushes with unbounded side-groups, showing
the asymptotic dependence $\langle r^2(t)\rangle\sim t^\alpha$ for the mean-square displacement along the base, with
$\alpha=1/D_w$. 
An important bonus result is  
the significant duration (more than $10^3$ steps) of the initial transient regime; see the inset of Fig.~4.
%This observation of course cannot be 
%foreseen  from a heuristic argument of the previous %section and suggests that for systems with finite %side-groups diffusion with transport exponent 
%(\ref{alpha_brush}) may be unobservable. 

%\enlargethispage{1\baselineskip}

Recall that  the matching argument leading to prediction  (\ref{alpha_brush}) is based on the assumption of a 
relative abruptness of the  crossover 
between the two asymptotic behaviors of the mean-square displacement in a system with finite side-groups. Our simulation confirms the validity of this assumption. Fig.~5 shows
$\langle r^2(t)\rangle$ for a Sierpinski brush with side-groups of length $2L$ for $L=100$
(and also for $L=50$ and $L=200$ in the inset). One observes that 
for short times
$t<t_c(L)$ the mean-square displacement follows the law 
$\langle r^2(t)\rangle\sim t^{\alpha}$ with $\alpha=1/D_w\approx 0.43$ as for the brush with infinite side-groups, while for
longer times $t>t_c(L)$ there occurs a transition to  the dependence 
$\langle r^2(t)\rangle\sim t^{2/D_w}=t^{0.86}$ as  for 
the isolated base. 
As can be seen in the inset, the characteristic transition 
time $t_c(L)$ increases with $L$ in a manner consistent  with theoretical expectation (\ref{tc}).

\begin{figure}
  \includegraphics[height=6.2cm]{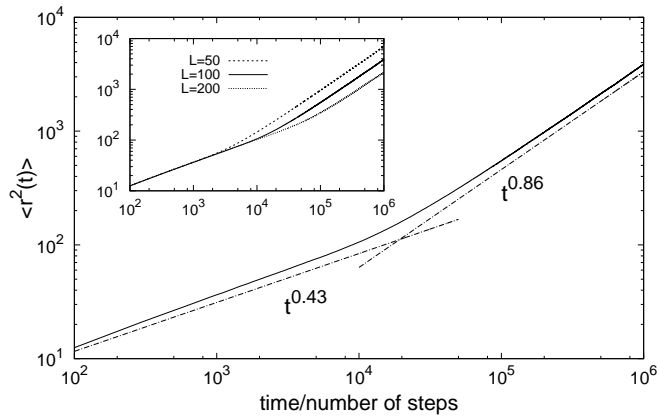}
  \caption{Simulation results for the mean-square displacement along 
  the base of the Sierpinski brush (depicted at Fig.~2) with side-groups of finite length $2L$.
  Solid line corresponds to  $L=100$, and dashed lines (in the inset) are for $L=50$ and $L=200$. The two dotted lines on the main plot show 
  short and long time asymptotic dependencies, 
  $t^\alpha$ with $\alpha=1/D_w\approx 0.43$ and $t^{2/D_w}$ with $2/D_w\approx 0.86$, 
  assumed by the matching argument of Sec. II. 
  }
\end{figure}

\section{Nonuniform systems: Cantor comb}
So far we have focused on uniform bundled structures for which 
side-groups are  anchored
to each site of the base, or for which they decorate the base  in a periodical manner. 
For such systems  
the fractal dimension $D_a$ of the set of anchor points
is the same as that of the base, $D_a=D$. As an example of a non-uniform system with $D_a<D$, let us consider a comb-like structure  composed of a 
one-dimensional base and side-groups, with side-groups attached to the base  at anchor points forming a Cantor set; see Fig.~6. For this system, which we call the Cantor comb, the set
of relevant dimensions is
\begin{eqnarray}
d=D=1,\quad d_w=D_w=2, \quad D_a=\ln 2/\ln 3\approx 0.63.
\end{eqnarray}

Compared to uniform systems, the presence of the additional dimension $D_a$ makes  a theoretical analysis of non-uniform bundled structures more complicated. To the best of our knowledge, so far 
there are no general results in the literature on transport  
properties of 
non-uniform bundled structures with an arbitrary set of dimensions.
A specific class of combs with a one-dimensional base and side-groups, $d=D=1$, and arbitrary $D_a<1$ (to which the Cantor comb belongs) was considered by Iomin~\cite{Iomin}. Using an analytical yet 
somewhat  ambiguous  approach, he suggested two possible expressions for the exponent $\alpha$ governing the mean-square displacement along the base $\langle x^2(t) \rangle\sim t^\alpha$. We found one of those expressions, namely
\begin{eqnarray}
\alpha=1-D_a/2,
\label{alpha_cantor}
\end{eqnarray}
to be consistent  with our simulation results for the Cantor comb.

Before outlining simulation details, let us stress that in contrast to the uniform systems result, (\ref{alpha_cantor}) presupposes (tacitly in Ref.~\cite{Iomin})
a special type of initial conditions. Namely,
initial sites of random walks  are assumed to be chosen randomly
among the anchor points, or 
within a finite distance from an anchor point.
In this case, after time $t$, when the walker  moved to the average distance $x(t)$ 
from an initial site, the average number of anchor points the walker 
encountered is estimated as $N_a(t)\sim x(t)^{D_a}$. For non-uniform systems, this estimation, assumed by the theory ~\cite{Iomin}, does not generally  hold for other choices for 
initial positions.  For example, if positions of initial sites
are chosen to be not correlated to that of anchor points then $N_a(t)$
would depend on  the system size $R$ and vanish in the limit 
$R\to\infty$ (since
the density of the anchor points  decreases with the system size $R$ as $1/R^{D-D_a}$). In that case, in the limit $R\to\infty$ side-groups do not affect diffusion at all, and instead of (\ref{alpha_cantor}) 
one would trivially expect to find  $\alpha=2/D_w$ ($\alpha=1$ for the Cantor comb),
i.e. the transport exponent for the isolated base. Our simulation 
of random walks on the Cantor comb with random initial positions, uncorrelated to positions of anchor points, indeed shows that sort of behavior.  
%Of course, for uniform systems ($D_a=D$), there is no such subtlety.   

Similar to the Sierpinski brush, it is convenient for simulation purposes to design the Cantor comb 
inductively from smaller to larger 
scales and to enumerate cells of the base with both decimal and 
ternary (base-3) integers~\cite{Plyukhin}.  As seen from Fig.~6,  this 
allows a simple test for an anchor cell: 
a cell is an  anchor cell (i.e. belongs to the Cantor set) 
if and only if  its ternary address 
has no  $1$-digits. 
The simulation results described below were obtained for the Cantor comb of generation 
$G_{30}$ with a base consisting of $3^{30}\sim 10^{14}$ cells. 
Further increasing or slightly
decreasing the generation order was found not to affect the results, which suggests that finite size effects were negligible in our simulation.

\begin{figure*}[t]
\includegraphics[height=5.7cm]{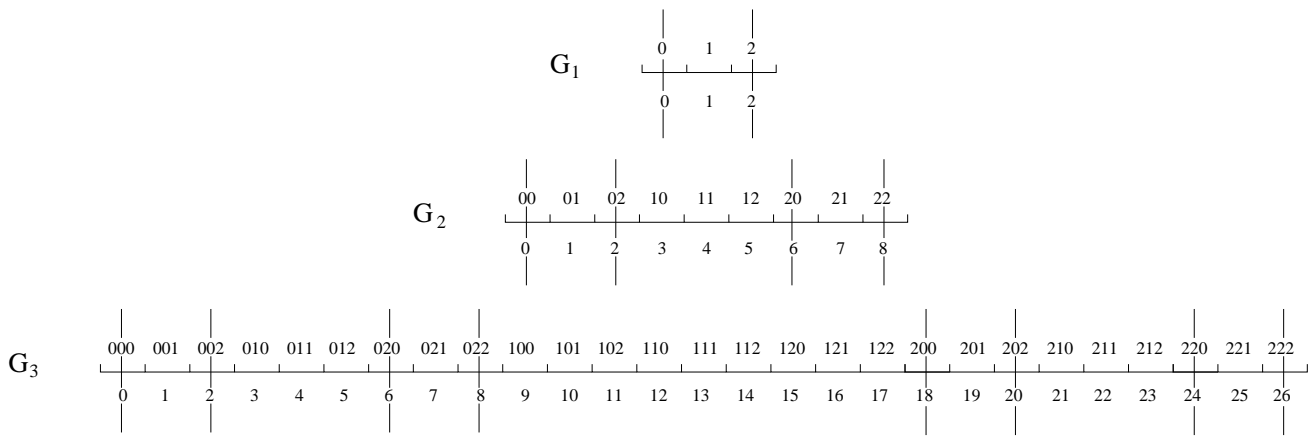}
\caption{
A recursive composition and cell enumeration of the Cantor comb. The first three generations, $G_1$, $G_2$, $G_3$, are shown. Below each base cell is its address in decimal notation, and above is 
the corresponding ternary (base-3) representation. Anchor cells, where side-groups are attached to the base, form the Cantor set and  have ternary addresses not containing the digit 1. 
}
\end{figure*}

\begin{figure}
  \includegraphics[height=6.2cm]{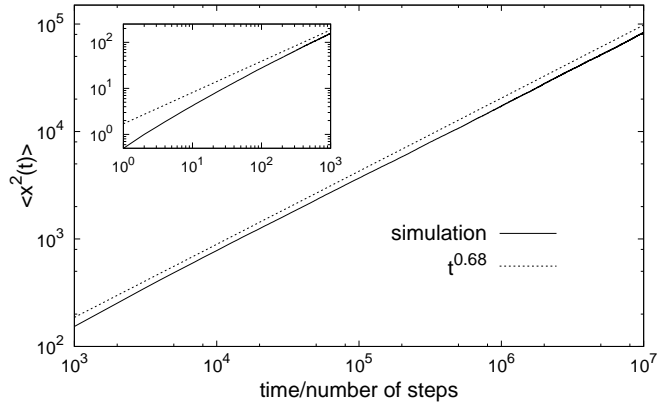}
  \caption{A log-log plot of the mean-square displacement 
$\langle x^2(t)\rangle$ along the base of the Cantor comb  (see Fig.~6)
of generation $G_{30}$
at long times (and at short times in the inset).
Solid lines show the simulation results averaged over about $10^5$
 walks, and dashed lines show the 
 theoretically expected dependence 
 $t^{\alpha}$
with the exponent given by Eq.(\ref{alpha_cantor}), $\alpha=1-D_a/2=1-\ln 2/(2\ln 3)\approx 0.68$.}
\end{figure}

\begin{figure}
  \includegraphics[height=6.2cm]{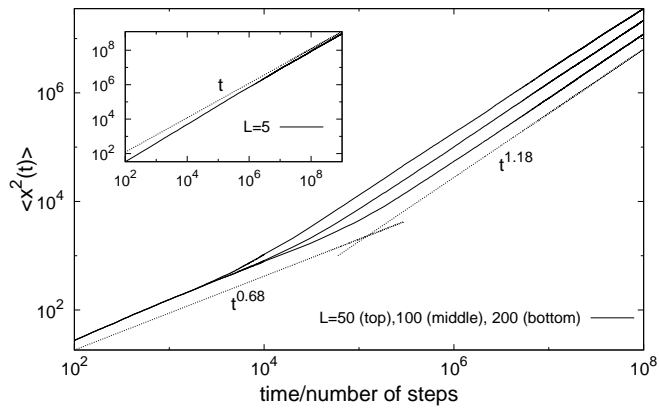}
  \caption{A log-log plot of the mean-square displacement 
$\langle x^2(t)\rangle$ 
%at long times (at short times in the inset) 
along the base of the Cantor comb of generation $G_{30}$
with finite side-groups of size $2L$.
The top solid line shows the simulation results 
for $L=50$, the middle one for $L=100$, and the bottom one for $L=200$.
For short time scales, the lines follow the sub-diffusion law $t^\alpha$ with 
$\alpha\approx 0.68$ (dashed line) as for the system with infinite side-groups. For longer time scales there occurs  a transition to the super-diffusive dependence $t^\beta$ with $\beta\approx 1.18$ (dashed line). 
For even longer time scales  the exponent $\beta$ decreases and tends toward one. The transition
to normal diffusion, $\beta\to 1$, is shown in the inset  
for a system with shorter side-groups,
$L=5$, for which the transition is noticeable  within the simulation time range.}
\end{figure}

For the Cantor comb with infinitely extended side-groups, the simulation results 
are shown in Fig.~7. The long-time asymptotic behavior of the mean-square 
displacement along the base
is well described by the sub-diffusion law $\langle x^2(t)\rangle \sim t^\alpha$ 
with the 
exponent  given by the theoretical prediction (\ref{alpha_cantor}), 
$\alpha=1-\ln 2/(2\ln 3)\approx 0.68$. 
Similar to the uniform Sierpinski brush, 
this behavior is preceded by a long (about $10^3$ steps) transient period of faster diffusion; see the inset of Fig.~7
%This suggests that for combs with teeth of large  
%but finite length the law 
%$\langle x^2(t)\rangle \sim t^{0.68}$ may be not observable.

\enlargethispage{1\baselineskip}

Now we consider random walks on the Cantor comb with side-groups of finite length $2L$.
Not only is this a question of a practical interest, but 
also along this line one may hope to extend for non-uniform systems 
the matching argument discussed  in Section II.
As for uniform systems, for short times,
while the diffusing particle does not reach the end points of side-groups,  
the mean-square displacement along the base of the Cantor comb 
is expected to follow the same sub-diffusion 
law as the comb with infinite side-groups,
i.e. $\langle x^2(t)\rangle \sim t^\alpha$ with $\alpha=1-D_a/2\approx 0.68$. Our simulation confirms this short-time asymptotic  behavior; see Fig.~8. 
However, in contrast to 
uniform combs, at longer times the initial sub-diffusion is followed  not by normal diffusion, as is characteristic of an isolated $1D$ base, but by a very long transient regime 
when $\langle x^2(t)\rangle$ increases with time faster than linearly. 
For this regime one can use, for limited time intervals,
a super-diffusion approximation   $\langle x^2(t)\rangle\sim t^\beta$ with $\beta>1$. However the exponent $\beta$ slowly decreases with time  tending to 
approach the value of one, as is characteristic of normal diffusion,  in the long time limit, $\lim_{t\to\infty} \beta(t)=1$. 
The duration of the super-diffusive regime increases with  the side-group's size $L$. 
For $L>50$ it is of more than $10^8$ steps and exceeds the time scale of our simulation; see Fig.~8.  The transition from super- to normal-diffusion occurs within the simulation time scale for systems with much shorter side-groups; the inset of Fig.~8 shows it for $L=5$.

Thus, diffusion along the base of the Cantor comb with finite side-groups shows not two, as for uniform systems, but three distinct regimes: initial sub-diffusion is followed up by  super-diffusion, which very slowly evolves into normal diffusion. 
We also found a similar behavior for a non-uniform version of the Sierpinski brush (see Fig.~2) where the anchor points form a Sierpinski gasket, but the base spans the entire two-dimensional lattice, 
$d=1, D_a=\ln 3/\ln 2, D=2$
(in contrast to the uniform Sierpinski brush, discussed in Section III, for which $D_a=D=\ln 3/\ln 2$).
We believe such a three-stage behavior is generic for non-uniform bundled systems with $D_a< D$. Clearly, the matching argument of Section II is not applicable here.

%\enlargethispage{2\baselineskip}

It is, of course, quite evident that a 
continuous transient from sub-diffusion to normal diffusion 
cannot be anything but super-diffusive. However, the very presence and significant duration of such a transient regime
may not be obvious.
One can intuitively interpret the origin of the super-diffusive regime in non-uniform bundled structures as 
follows: On time scales much longer than $t(L)\sim L^{d_w}$ 
the role of side-groups of size $L$ is to
reduce the effective jumping rate  of random walks along the base. As time progresses, a diffusing particle explores regions of larger spatial scale $x$ of the base, and ``sees'' the density of side-group sites decreasing as $1/x^{D-D_a}$. As a result, 
the effective jumping rate and diffusion coefficient increase with $x$ which, 
as is well known~\cite{Risken}, may result in super-diffusive transport. On the other hand, 
in the long time limit, when the diffusing particle explores a vast region for which the
fraction of side-groups  is negligible compared to the number of sites of the base, diffusion is expected to follow asymptotically the law 
$\langle x^2(t)\rangle \sim t^{2/D_w}$  (that is, $\langle x^2(t)\rangle \sim t$ for combs) 
as for the isolated base.

\section{Conclusion}

We have studied random walks on two groups of comb- and brush-like graphs. 
The first group is comprised of systems uniformly decorated with side-groups,
for which the fractal dimensions of the base and the set of anchor points are equal, $D=D_a$.
The second group is that of  non-uniform graphs for which  
anchor points form a proper 
subset of the base with dimension smaller than that of the base, $D_a<D$.

For uniform graphs we extended the qualitative argument by Forte et al.,
originally developed for comb-like systems ($D=1$), to brushes with 
arbitrary base dimension $D$. The result is expression (\ref{alpha_brush}) for the exponent $\alpha$, describing sub-diffusive transport along the base of the system with unbounded side-groups.  
We verified this expression in numerical simulations for the uniform Sierpinski brush with $D=D_a=\ln 3/\ln 2$.
The simulation also shows that the regime of stationary sub-diffusion with 
$\langle r^2(t)\rangle\sim t^\alpha$ is preceded by a transient regime of significant duration (about $10^3$ steps) and therefore the former 
may not be observable in systems with finite and relatively short side-groups.

For  uniform graphs with finite but sufficiently long side-groups,
simulation shows
that the mean-square displacement along the base 
experiences a transition from  the
time dependence $t^\alpha$
(as in a system with infinite side-groups) at shorter time scales
to  $t^{2/D_w}$ (as for the isolated base) at longer time scales.
Simulations confirm the  abruptness of this transition, which is 
tacitly implied in the argument by Forte et al.
%The occurrence time of the transition increases with the side-group size $L$, 
%but the transition's duration does not. 

In contrast, for nonuniform graphs like the Cantor comb ($D=1$, $D_a=\ln 2/\ln 3$) with side-groups  of finite length, the transition to the time dependence  $\langle x^2(t)\rangle \sim t^{2/D_w}$, characteristic for an isolated base, occurs only after a very long regime of super-diffusive transport.
While the origin of the super-diffusive regime is intuitively clear, its theoretical description remains an open problem.

For non-uniform fractal combs with infinite side-groups, 
our simulation supported that the asymptotic mean-square displacement along the base $\langle x^2(t)\rangle\sim t^\alpha$ is characterized by an exponent of the form
$\alpha=1-D_a/2$, as suggested by Iomin~\cite{Iomin}. It still remains to be seen how to generalize 
this result for non-uniform bundled structures  with $d, D>1$. As a reference point for a future theory we  evaluated $\alpha$ for a non-uniform brush (briefly mentioned in the previous section) with
one-dimensional side-groups, a two-dimensional base, and the anchor set given by 
the Sierpinski gasket,
\begin{eqnarray}
d=1,\qquad d_w=2, \qquad D=D_w=2,\qquad D_a=\ln 3/\ln 2.
\end{eqnarray}
The structure is illustrated by the same Fig.~2 as for the Sierpinski brush discussed in Section II, but now a walker is allowed to step on every cell of the 2D lattice. For this non-uniformly decorated brush ($D_a<D$) we observed transport properties qualitatively similar to that of the Cantor comb, and for the mean-square exponent we obtained the approximate value $\alpha\approx 0.72$.

The simulation schemes described in the paper can be readily implemented for biased random walks
to evaluate a directional drift, i.e. the mean displacement along the base $\langle r (t)\rangle_f$  induced by a weak external constant field $f$. The latter is modeled by replacing the unbiased jumping probability $p$ to $p\pm f$ for jumps in the direction  along/against the field.
In previous works it was found that for uniform combs 
the field induced drift $\langle r(t)\rangle_f$  
and  the mean-square displacement for unbiased ($f=0$) 
random walks $\langle r^2(t)\rangle$ satisfy the (generalized) Einstein relation,
both  increasing  with time according to the same law~\cite{Havlin_paper,Havlin_book,Forte,Vulpiani}
\begin{eqnarray}
\langle r (t)\rangle_f\sim \langle r^2(t)\rangle\sim t^\alpha.
\label{ER}
\end{eqnarray}
Our simulations showed the validity of this relation 
also for uniform brushes like the Sierpinski brush and for non-uniform  structures like the Cantor comb. We found  
relation (\ref{ER}) valid for systems with finite and infinite side-groups, and not only for for asymptotic but for 
transient (in particular, super-diffusive) regimes as well. However, and remarkably, for the Cantor comb we found the deviation from the Einstein  relation (\ref{ER}) to be noticeable already for very weak fields $f>10^{-3}$.

\end{document}